\documentclass[12pt]{article}
\pdfoutput=1
\usepackage{putex}
\usepackage{amsmath}
\usepackage{amssymb}

\usepackage{graphicx}
\usepackage[lofdepth,lotdepth]{subfig}
\usepackage{epstopdf}
\usepackage{enumerate}
\usepackage{cite}
\usepackage{tensor}
\usepackage{slashed}
\usepackage{feynmf}
\usepackage{hyperref}
\usepackage{bbold}
\usepackage{dsfont}
\usepackage{mathrsfs}
\usepackage{caption}
\usepackage{comment}

\usepackage{youngtab}
\usepackage{amsfonts}
\usepackage{braket}

\newcommand{\hb}{\overline{h}}

\renewcommand{\theequation}{\thesection.\arabic{equation}}

\usepackage[utf8x]{inputenc}
\usepackage{braket}
\usepackage{titlesec}
\usepackage{fixltx2e}
\usepackage{cleveref}
\hypersetup{breaklinks}

\usepackage[usenames,dvipsnames,svgnames,table]{xcolor}

\def\Oc{{\cal O}}

\def\p{\partial}

\newcommand {\be} {\begin {equation}}
\newcommand {\ee} {\end {equation}}
\newcommand{\bea}{\begin{eqnarray}}
\newcommand{\eea}{\end{eqnarray}}
\newcommand{\dDisc}{\text{dDisc}}
\newcommand{\F}{{}_2F_1}
\newcommand\numberthis{\addtocounter{equation}{1}\tag{\theequation}}

\def\zb{\overline{z}}

\def\rt{\rightarrow}

\def\Lt{\tilde{L}}
\def\nb{\overline{n}}
\def\pb{\overline{\partial}}
\def\Oc{\mathcal{O}}
\def\Hb{\overline{H}}
\def\Ic{\mathcal{I}}

 \newmuskip\pFqmuskip

\newcommand*\pFq[6][8]{%
  \begingroup 
  \pFqmuskip=#1mu\relax
  \mathcode`\,=\string"8000
  \begingroup\lccode`\~=`\,
  \lowercase{\endgroup\let~}\pFqcomma
  {}_{#2}F_{#3}{\left[\genfrac..{0pt}{}{#4}{#5};#6\right]}%
  \endgroup
}
\newcommand{\pFqcomma}{\mskip\pFqmuskip}

\makeatletter
\renewcommand{\@maketitle}{
\newpage
 \begin{center}%
  {\large\bfseries \@title \par}%
 \end{center}%
 \par} \makeatother

\numberwithin{equation}{section}

\titleformat*{\section}{\large\bfseries}

\begin{document}

\institution{UCLA}{ \quad\quad\quad\quad\quad\quad\quad\quad\quad Mani L. Bhaumik Institute for Theoretical Physics, \cr Department of Physics and Astronomy, University of California, Los Angeles, CA 90095, USA}

\title{Late time Wilson lines
}

\authors{Per Kraus, Allic Sivaramakrishnan, and  River Snively}

\abstract{In the AdS$_3$/CFT$_2$ correspondence, physical interest attaches to understanding Virasoro conformal blocks at large central charge and in a kinematical regime of large Lorentzian time separation, $t\sim c$. However, almost no analytical information about this regime is presently available.    By employing the Wilson line representation  we derive new results on conformal blocks at late times, effectively resumming all dependence on $t/c$.  This is achieved in the context of ``light-light" blocks, as opposed to the richer, but much less tractable, ``heavy-light" blocks.    The results exhibit an initial decay, followed by erratic behavior and recurrences.    We also connect this result to gravitational contributions to anomalous dimensions of double trace operators by using the Lorentzian inversion formula to extract the latter. Inverting the stress tensor block provides a pedagogical example of inversion formula machinery.
    }

\date{}

\maketitle
\setcounter{tocdepth}{2}
\tableofcontents

\section{Introduction}

Although the AdS/CFT correspondence establishes that Einstein gravity
can coexist with black holes evolving unitarily into Hawking radiation,
much remains mysterious about how precisely this comes about in the
language of bulk gravitational physics.  See for example \cite{Harlow:2014yka,Polchinski:2016hrw} for reviews.  Some of the most obvious
questions of physical relevance, such as the experience of an observer
falling through the horizon, are not easily posed, much less answered,
in terms of boundary CFT correlators.   Maldacena suggested focusing
on a CFT quantity that is well defined and does address some of  the
puzzles involving black holes \cite{Maldacena01}. Namely, a boundary two-point correlation
function at large Lorentzian time separation decays exponentially to
zero as $e^{-at/\beta}$ in the semi-classical bulk approximation, yet by
unitarity has a long time average bounded below by $e^{-bS}$.  Here $a$
and $b$ are numerical factors, $\beta $ is the inverse temperature and
$S$ is the entropy. The implication is that the bulk semi-classical
approximation must fail at late times, $t\sim \beta S$.   Essentially, a
reliable computation at this time scale needs to incorporate that the
black hole has a discrete energy spectrum rather than the continuous
spectrum that arises semi-classically.

This paper is focussed on the AdS$_3$/CFT$_2$ correspondence, and in particular on the
CFT$_2$ side.  The bulk semi-classical limit corresponds to large
central charge, $c\gg 1$, with kinematical factors held fixed.  The
issue at hand is that the relevant late time correlators have time
separations $t \sim c$, invalidating this approximation.  Thus we need
to develop new analytical tools for understanding correlation functions
in this regime.   Here we report on  progress in this direction,
although we will  make only  indirect contact with the deep quantum gravity
questions that form the underlying motivation for this work.

Universal aspects of correlation functions in 2d CFT are captured by the
Virasoro conformal blocks (e.g, \cite{Hartman13,FitzpatrickKW14,Hijano:2015qja}), and so interest attaches to understanding
these at large Lorentzian times.  Here we are considering time to be
defined on the Lorentzian cylinder. While a good deal is known
analytically about these conformal blocks at large $c$, this does not so
far extend to the regime of cross ratio space corresponding to large
Lorentzian time.  For a numerical study, see \cite{ChenHKL17}. Our approach
is via the Wilson representation of conformal blocks, as developed in
\cite{Bhatta:2016hpz,Besken:2016ooo,FitzpatrickKLW16,BeskenHK17,HikidaU117,HikidaU217,HikidaU18,Besken:2018zro}.
Here, the conformal block corresponding to $\Oc \Oc \rt {\rm
stress~tensors} \rt \Oc' \Oc'$ is expressed as $\langle h,
h'|Pe^{\int_{z_1}^{z_2} (L_1 + {6\over c}T(z)L_1)dz}|h,h'\rangle$, where
$T(z)$ is the CFT stress tensor.  Details of this construction are
reviewed in the next section.   The conformal block is a contribution to
the correlator that can be written as $\langle \Oc'(\infty) \Oc(t_1) \Oc(t_2)
\Oc'(-\infty)\rangle$ on the cylinder, or as $\langle \Oc'(\infty) \Oc(z_1) \Oc(z_2)
\Oc'(0)\rangle$ on the plane.  As a function of $z_2$, this Wilson line has a
branch cut running between the locations of the two $\Oc'$ operators,
and going to late Lorentzian time separation between the $\Oc$ operators
corresponds to taking $z_2$ to wind many times around the branch
point.     In terms of making contact with black hole physics, the
relevant regime is one in which $c\rt \infty$, with $h, h'/c$ and $N/c$
fixed, where $N$ denotes the number of windings.  The conditions $h\sim
c^0$ and $h'\sim c$ are desired so that the correlator represents, in
bulk language, a light particle probing a heavy state in the black hole
regime.     Evaluating the Wilson line in this regime is very challenging
since we need to evaluate the infinite sum of nested integrals implied
by the path ordered exponential.   We therefore consider a simpler
regime in which $h'$ is held fixed as $c\rt \infty$.    Here the late
time behavior is much more tractable but still nontrivial.   Our main
result is to demonstrate how the Wilson line  efficiently captures this
regime, essentially resumming all terms in the $t/c$ expansion.

The answer turns out to be very simple to describe.   First note that the $c\rt \infty$
limit with everything else held fixed, including $t$, corresponds to the
free field limit in the bulk, and the Wilson line reduces to a product
of two-point functions, $\langle \Oc \Oc\rangle \langle
\Oc'\Oc'\rangle$.  Equivalently,  in terms of conformal blocks appearing
in the expansion $\Oc \Oc \rt \Oc_p \rt \Oc' \Oc'$, only the identity
operator appears.  Writing this result as a Fourier sum on the cylinder
gives an expression of the form $\sum_n A_n e^{-iE_n t}$, where $E_n  =
h+h'+ n$, and the $A_n$ are essentially the OPE coefficients appearing
in the block expansion $\Oc \Oc' \rt [\Oc \Oc']_n \rt \Oc \Oc'$, where
$[\Oc \Oc']_n$ are the double trace operators of dimension $h+h'+n$.  If
we instead keep all dependence on $t/c$ as $c\rt \infty$, we show that
the only modification to this result is that $E_n \rt E_n +{\gamma_n\over 2}$ with $\gamma_n= -{12\over c}
\big[C_2(h+h'+n)-C_2(h)-C_2(h')   \big] $, where $C_2(h) = h(h-1)$ is
the SL(2) quadratic Casimir.  This result can be thought of as coming from  exponentiating the global stress tensor block in the late time regime.   Since the $E_n$ are no longer integers in
general, this result has a much more complicated time dependence, as we
illustrate with some representative plots.  After introducing a
regulator to smooth out lightcone singularities, the result exhibits an
initial decay at early time followed by an erratic late time behavior,
including recurrences.   This is the sort of behavior one hopes to see
to address black hole physics, but we emphasize again that we cannot
make any direct connection here, both because we are only considering
low dimension operators and because we are only considering a Virasoro
block and not a full correlator.   But we hope that this does provide a
useful warmup example involving late time resummation.

To better understand the shift $E_n \rt E_n +{\gamma_n\over 2} $  appearing in the Wilson line result, we compare to anomalous dimensions arising from tree level  graviton exchange Witten diagrams in AdS$_3$.
Expanding such a diagram in the crossed channel
yields the anomalous dimensions (and OPE coefficients) of the double
trace operators of schematic form $\Oc \p^n \pb^{\nb} \Oc'$. Extending recent advances in the analytic bootstrap \cite{LargeSpinPerturbationTheory, Simmons-Duffin:2016wlq}, the Lorentzian inversion formula \cite{OPEInversion} provides a particularly efficient way to compute such quantities.
In this approach, inverting a tree-level Witten diagram boils down to inverting a conformal block.
The process of inverting a block is a starting point for using powerful
inversion-formula technology to investigate higher-loop effects \cite{AldayC17, AldayHL2017}, and
general results have appeared recently in \cite{Sleight:2018ryu,CardonaS18,LiuPRS18}.
As the inversion formula is emerging as a remarkably useful tool for studying AdS/CFT,
we aim to provide a worked example that displays the nuts and bolts of inverting blocks in a way accessible
to those unfamiliar with  analytic bootstrap machinery.

The starting point corresponds to inverting the identity exchange, which is the simplest case.
By then including also the stress tensor block, our case of interest, we find
anomalous dimensions $\gamma_{n,\nb} = -{12\over c}  \big[C_2(h+h'+{\rm
min}(n,\nb) )-C_2(h)-C_2(h')   \big]$.  The similarity with the Wilson line
result is evident, though note that it is $ {\rm min}(n,\nb)$ that
appears in the anomalous dimension (in terms of the twist and spin of
the double trace operators, this says that the anomalous dimension
depends solely on the twist). While they appear as corrections to $E_n$, the $\gamma_n$ that govern the behavior of the Wilson line cannot immediately be identified as  bonafide  anomalous dimensions  because $\gamma_n$ arise from a single conformal block rather than a full correlator; the latter includes also exchanges of double trace operators, leading to the appearance of $\gamma_{n,\nb}$ as the anomalous dimensions.

However, the correspondence between $\gamma_n$ and the proper anomalous dimensions $\gamma_{n,\nb}$ can be understood by considering  the lightcone limit, which is $\zb \rt 1$ in our setup.  This limit projects out exchanges with nonzero $\overline{h}$, leaving just the holomorphic component of the stress tensor, which is what the Wilson line captures.  Further, in the crossed channel expansion this limit corresponds to the large $\overline{h}$ regime.  Finally, on general grounds we know that anomalous dimensions of double trace operators due to graviton exchange in AdS$_3$ are spin independent; this follows in the CFT from analyticity in spin combined with the known asymptotic behavior.  Putting these facts together, we see that $\gamma_n$ with $n$ interpreted as twist are the anomalous dimensions of the family of double-twist operators with $\nb>n$, thus explaining the agreement between the Wilson line and the results obtained from the full correlator.

The remainder of this paper is organized as follows.  In section 2 we
review the construction of the Wilson line, and in section 3 we discuss
how to evaluate it.   To illustrate its use we first rederive the known
result for the Virasoro block in the limit $c\rt \infty$ with ${h\over
c}, {h'\over c} \rt 0$, but ${hh'\over c}$ and $t$ fixed.  In this limit the
Virasoro block is the exponential of the global stress tensor block  \cite{FitzpatrickKW14}, a
result which is obtained from the Wilson line with minimal labor.  We then
turn to the main case of interest involving late times, and again show
how the Wilson line deals with this efficiently.   In section 4 we
discuss in some pedagogical detail the computation of anomalous
dimensions using the Lorentzian inversion formula.    We conclude with
some comments in section 5.

\section{The Virasoro Wilson line}

In this section we recall the basic construction of the Wilson line, and how it provides a  representation of Virasoro conformal blocks that admits a convenient expansion at large central charge. See \cite{Bhatta:2016hpz,Besken:2016ooo,FitzpatrickKLW16,BeskenHK17,HikidaU117,HikidaU217,HikidaU18,Besken:2018zro} for more background and previous results.

Consider a primary operator $\Oc(x)$.  We can use the OPE to expand $\Oc(x_1)\Oc(x_2)$ in terms of local operators at some point $x_3$.   Organizing the expansion in representations of the Virasoro algebra corresponds to collecting terms that differ only in the number of stress tensors that appear.  Schematically,
\bea\label{aj}
\Oc(x_1) \Oc(x_2) =  [1+ T + TT+ \ldots] +\sum_i C_{\Oc \Oc \Oc_i} [\Oc_i + \Oc_i T  + \Oc_i TT+ \ldots]
\eea
where we have suppressed numerical coefficients (which depend on the central charge $c$) and the dependence on coordinates and derivatives. Except for the OPE coefficients $C_{\Oc \Oc \Oc_i}$ and the spectrum, everything is fixed by the Virasoro algebra. Since the full symmetry algebra is two copies of Virasoro associated to $T(z)$ and $\overline{T}(\zb)$, each term above is really a product of a $T$ piece and $\overline{T}$ piece, but we henceforth focus on the $T$ piece alone.     The first term in the expansion (\ref{aj})  is the Virasoro vacuum OPE  block. The Wilson line is conjectured to provide a representation of the Virasoro vacuum OPE block \cite{FitzpatrickKLW16},
\bea\label{ak}
\Oc(x_1) \Oc(x_2) =  \langle h;{\rm out}| P e^{\int_C a(z) }|h;{\rm in}\rangle +[{\rm non-pure ~stress~tensor~terms}]
\eea
with
\bea\label{al}   a(z)  =  (L_1 + {6\over c}T(z)L_{-1})dz~.
\eea

Let us explain the ingredients in this construction.  $P$ denotes path ordering along the contour $C$ that runs from $z_1$ to $z_2$, with operators at later points on the contour  moved to the left.  The states $|h;{\rm in}\rangle$ and $|h;{\rm out}\rangle$ lie in representations of the SL(2) algebra with generators $L_{-1,0,1}$ which obey
\bea\label{am} [L_m,L_n]=(m-n)L_{m+n}~.
\eea
The inner product is defined such that $L_n^{\dagger}  = L_{-n}$.
The states obey
\bea\label{an}
&&L_{-1}|h;{\rm in}\rangle =0~,\quad L_0 |h;{\rm in}\rangle =-h|h;{\rm in}\rangle~, \cr
&& L_{1}|h;{\rm out}\rangle =0~,\quad L_0 |h;{\rm out}\rangle =h|h;{\rm out}\rangle~.
\eea
In (\ref{al}) $T(z)$ is a stress tensor operator (as opposed to a classical function); unlike in (\ref{aj}) these stress tensors are smeared over the contour running between $z_{1,2}$.   The stress tensor $T(z)$ does not talk to the SL(2) generators; in particular, $L_n$ do {\em not} appear in the mode expansion of $T(z)$.

A few more comments are in order before we justify the above relation between the Wilson line and the Virasoro vacuum OPE block.  First, the Wilson line is a divergent object since the integral involves colliding stress tensors, whose OPE is singular.  These divergences can be renormalized by including a multiplicative renormalization factor in front of the Wilson as well as introducing a vertex renormalization:  ${6\over c}T(z) \rightarrow {6\alpha \over c}T(z)$.  Finiteness and Ward identities uniquely fix the divergent and finite parts that appear in a suitable dimensional regularization scheme \cite{HikidaU18,Besken:2018zro}.  Second, the states that appear in the definition of the Wilson line involve a quantity $h$.   In the large $c$ limit, $h$ coincides with the scaling dimension of the operator $\Oc(x)$, but at finite $c$ they differ in a known way \cite{Besken:2016ooo}.   These two renormalization issues will not be relevant here, since we work in the large $c$ limit, so we do not dwell on them further.

To streamline notation a bit, we henceforth write
\bea\label{ao}  |h;{\rm in}\rangle = |-h\rangle~,\quad  \langle h;{\rm out}| = \langle h|.
\eea
 The appearance of $\pm h$ denotes the $L_0$ eigenvalue.

To fully establish the equivalence of the Wilson line and the Virasoro vacuum block we should prove  that correlators involving any number of stress tensor insertions are correctly reproduced,
\begin{align*}
\label{ap}
&\langle 0_{\rm CFT}| \Oc(x_1) \Oc(x_2) T(z_3) \ldots T(z_n)|0_{\rm CFT} \rangle
\\
&\quad\quad\quad=  \langle 0_{\rm CFT}| \langle h|Pe^{\int_{z_1}^{z_2} a} |-h\rangle ~ T(z_3) \ldots T(z_n)|0_{\rm CFT} \rangle ~.
\numberthis
\end{align*}
Here we are taking the CFT vacuum expectation value on both sides, in addition to computing the SL(2) matrix element on the right.  While there is good evidence for the claim, as established in the references cited above, it has not been proven in full generality, and in particular the renormalization issues remain to be fully worked out to all orders.

The logic behind the association of the Wilson line with the Virasoro vacuum block stems from the relation between SL(2) transformations and conformal transformations.  That is, the Wilson line is built purely out of stress tensors yet enjoys the conformal transformation properties of the bilocal object $\Oc(x_1)\Oc(x_2)$.  Proving this in general is equivalent to establishing (\ref{ap}), which  we have said requires a careful renormalization treatment.   Instead,  let us consider something simpler.  Let the CFT state be such that $T(z)$ has a classical expectation value at large $c$, $\langle T(z)\rangle\sim c$.  In this regime the $T(z)$ operator appearing in the Wilson line can be replaced by its expectation value, which we continue to denote by $T(z)$.

To establish the transformation properties of the Wilson line in this classical limit,  consider the $z$-dependent SL(2) group element
\bea\label{aq}
U(z) = e^{\lambda_1(z)L_1} e^{\lambda_0(z)L_0} e^{\lambda_{-1}(z) L_{-1} }
\eea
with
\bea\label{ar}
\lambda_1 = z-f(z)~,\quad \lambda_0(z) =- \ln [f'(z)]~,\quad \lambda_{-1}(z) = -{f''(z)\over 2f'(z)}~.
\eea
Wilson lines in gauge theories transform in a well known way under gauge transformations, which in our case amounts to
\bea\label{as}
U^{-1}(z_2) P e^{\int_{z_1}^{z_2} a_U(z)} U(z_1) = P e^{\int_{z_1}^{z_2} a_U(z)}
\eea
with
\bea\label{at} a_U(z) = U^{-1}(z) a(z)U(z)-U^{-1}(z)dU(z)~.
\eea
If we choose $a(z)= L_1 dz$, corresponding to vanishing stress tensor, we find $a_U(z)= (L_1 +{6\over c}T(z)L_{-1})dz$ with
\bea\label{au} T(z) =   {c\over 12} S_f(z)~,\quad S_f(z) = {f'''(z)\over f'(z)}-{3\over 2} \left( f''(z)\over f'(z)\right)^2~.
\eea
$S_f(z)$ is the Schwarzian derivative.   If we now take the SL(2) matrix element using (\ref{an}) we find
\bea\label{av}
\langle h| P e^{\int_{z_1}^{z_2} (L_1+ {6\over c}T(z)L_{-1})dz}|-h\rangle =  [f'(z_2)f'(z_1)]^h \langle h| P e^{\int_{f(z_1)}^{f(z_2)} L_1dz}|-h\rangle.   \eea
%
%
%
Since $ \langle h| P e^{\int_{z_1}^{z_2} L_1dz}|-h\rangle = (z_2-z_1)^{-2h}$ (see below), we find that (\ref{av}) reads
\bea\label{ax}  \langle h| P e^{\int_{z_1}^{z_2} (L_1+ {6\over c}T(z)L_{-1})dz}|-h\rangle =  { [f'(z_2) f'(z_1)]^h \over [ f(z_2)-f(z_1)]^{2h}}~.
\eea
This makes perfect sense as it says that if we generate a stress tensor by performing a conformal transformation $z\rt f(z)$, the Wilson line result takes the form of a primary two-point function transformed by $f(z)$.  Again, we stress that these statements have been established in the classical limit.

If we take  the expectation value of  the Virasoro vacuum OPE block in a CFT primary state $|h_{\rm CFT}\rangle$ we obtain the Virasoro vacuum block in the channel $\Oc_h \Oc_h \rt {\rm stress~tensors} \rt \Oc_{h_{\rm CFT}} \Oc_{h_{\rm CFT}}$,
\bea\label{ay} V_{h,h_{\rm CFT}}(z_1,z_2)=  \langle h; h_{CFT}| P e^{\int_{z_1}^{z_2} a(z) } |-h;h_{\rm CFT}\rangle~,
\eea
where the states are defined  in the tensor product of SL(2) times Virasoro. The connection is $  a(z)  =  (L_1 + {6\over c}T(z)L_{-1})dz$, and evaluating the right hand side of (\ref{ay}) means to expand the exponential and then use the fact that all stress tensor correlators $ \langle h_{CFT}| T(z_3) \ldots T(z_n) |h_{\rm CFT}\rangle$ are fully determined by conformal symmetry.  In terms of four-point functions, the conjectured relation between the Wilson line and the Virasoro amount block is
\bea\label{az}  V_{h,h_{\rm CFT}}(z_1,z_2) = \langle \Oc_{h_{\rm CFT}}(0)\big[ \Oc_h(z_1) \Oc_h(z_2)\big]_{\rm vac} \Oc_{h_{\rm CFT}}(\infty)\rangle~,
\eea
where $\big[ \Oc_h(z_1) \Oc_h(z_2)\big]_{\rm vac}$ means that we take the OPE as in (\ref{aj}) and keep only the stress tensor terms.  We could use conformal symmetry to send $z_1$ to a specified location and identify $z_2$  with the conformal cross ratio.

No usable closed form expression for the Virasoro vacuum block is known; see \cite{Perlmutter:2015iya} for a useful review.  From our point of view, the technical challenge lies in evaluating the nested integrals of stress tensor correlators that arise upon expanding the path ordered exponential.

 Now let us discuss more about the evaluation of the SL(2) matrix elements, for instance $\langle h| e^{L_1 z}|-h\rangle$. One way to proceed is to first take $h=-j$, where $2j$ is a non-negative integer.  In this case, we have a finite dimensional (non-unitary) representation realized by $(2j+1)\times (2j+1)$ matrices. One can then work out results for arbitrary $j$ and then analytically continue $h=-j$ to positive values; this last step of course requires some knowledge of the analytic structure in the complex $h$ plane.  So, for example, since $L_1$ lowers the $L_0$ eigenvalue by one unit, it is immediately clear that $\langle -j| e^{L_1 z}|j\rangle = z^{2j}$ up to normalization, and then analytic continuation yields $z^{-2h}$, which is the desired form of the two-point function.  At least in $1/c$ perturbation theory, all computations can be done in this manner, and in fact this is a very efficient way to proceed.

Alternatively, one can work directly with unitary representations, for example by realizing SL(2) in terms of functions of the complex variable $u$ defined on the unit disk.  We write
\bea\label{ba} L_1 = \partial_u~,\quad L_0 = u\partial_u +h~,\quad L_{-1}=u^2 \partial_u+2hu~.
\eea
The inner product between functions $f(u)$ and $g(u)$ is defined as an integral over the unit disk,
\bea\label{bb}  \langle f|g\rangle = \int_D  {d^2u\over  (1-u\overline{u})^{2-2h}} \overline{f(u)}g(u)~.
\eea
This is defined to respect the relations $L_n^\dagger =L_{-n}$.  The states appearing in the Wilson line correspond to functions,
\bea\label{bc} |h\rangle \rightarrow 1~,\quad   |-h\rangle \rightarrow u^{-2h}~.
\eea
For generic $h$ these two states are not in a common irreducible SL(2) representation, but this fact poses no problem in the construction since the inner product has been defined for all functions on the unit disk.
We then have
\bea\label{bd} \langle h| e^{L_1 z}|-h\rangle  =  \int_D  {d^2u\over  (1-u\overline{u})^{2-2h}} (u+z)^{-2h} \propto z^{-2h} ~,
\eea
where the $z$ dependence is immediately fixed by rotational symmetry on the disk.

To close this section, we should also mention that the most intuitive explanation for the form of the Wilson line comes from thinking about the AdS$_3$/CFT$_2$ correspondence, and in particular from the fact that AdS$_3$ gravity is equivalent to SL(2)$\times $SL(2) Chern-Simons theory; actually, these constructions originated in higher spin extensions \cite{deBoer:2013vca,Ammon:2013hba,deBoer:2014sna,Hegde:2015dqh}.  Asymptotically AdS$_3$ solutions of Einstein's equations (which are all locally AdS$_3$ since there are no dynamical degrees of freedom)  are recast as flat connections.   The connection $A(z) = (e^\rho L_1 +{6\over c}T(z)e^{-\rho} L_{-1})dz$ arises in this way, where $T(z)$ is identified via the holographic dictionary as being the boundary stress tensor. $\rho$ is a radial coordinate, which can be ``gauged away" to obtain the reduced connection $a(z)$ appearing in our Wilson line.  In this way, one sees that the Wilson line is simply a standard Wilson line for the bulk Chern-Simons connection, with endpoints on the AdS$_3$ boundary where dual CFT operators are located. In this context, the notion of taking $T(z)$ to be an operator corresponds to performing the Chern-Simons path integral, rather than restricting to  a fixed classical background.

\section{Evaluating the Wilson line}

Our task is to evaluate the Wilson line
\bea\label{aya} V_{h,h_{\rm CFT}}(z_1,z_2)=  \langle h; h_{CFT}| P e^{\int_{z_1}^{z_2} (L_1 +{6\over c}T(z) L_{-1})dz} |-h;h_{\rm CFT}\rangle~.
\eea
We now do some additional rewriting and processing to facilitate computation.
First, we henceforth write
\bea\label{bh}  h_{\rm CFT} = h'~.
\eea
Next, since
the Wilson line involves  $P e^{\int_{z_1}^{z_2} (L_1 +{6\over c}T(z) L_{-1})dz}$,  we can think of $z$ as time and $L_1 +{6\over c}T(z) L_{-1}$ as a time dependent Hamiltonian.  As in many applications,  it is useful to think of $L_1$ as a free Hamiltonian and ${6\over c}T(z) L_{-1}$ as an interaction.  Then, we can implement the same steps one takes to pass to the interaction representation by writing the identity
\bea\label{be}
P e^{\int_{z_1}^{z_2} (L_1 +{6\over c}T(z) L_{-1})dz} =  e^{L_1 z_2}  P e^{{6\over c}\int_{z_1}^{z_2} H_I(z)dz}e^{-L_1 z_1} ~,
\eea
with
\bea\label{bf} H_I(z) =  \big(L_{-1}-2zL_0+z^2L_1\big)T(z)~.
\eea
We then have to evaluate
\bea\label{bg} V_{h,h'}(z_1,z_2)= \langle h,h'|   e^{L_1 z_2}  P e^{{6\over c}\int_{z_1}^{z_2} H_I(z)dz}e^{-L_1 z_1} | -h,h'\rangle.
\eea
Next, it is useful to use the identity
\bea\label{bda} \langle h |e^{L_1 z_2} = z_2^{-2h}\langle -h| e^{-{1\over z_2} L_{-1}} . \eea
This is easily derived using the representation of the SL(2) generators given in (\ref{ba}), where it becomes an equality of functions on the unit disk.  This identity gives
\bea\label{bga} V_{h,h'}(z_1,z_2)= z_2^{-2h} \langle -h,h'| e^{-{1\over z_2} L_{-1}}    P e^{{6\over c}\int_{z_1}^{z_2} H_I(z)dz}e^{-L_1 z_1} | -h,h'\rangle .
\eea
Finally, to exhibit the symmetry between $h$ and $h'$ we redefine the SL(2) generators as
\bea\label{bq} \Lt_{-1}=-L_1~,\quad \Lt_0 = - L_0~,\quad \Lt_1 = -L_{-1}~.
\eea
This preserves the SL(2) algebra, $[\Lt_m,\Lt_n]=(m-n)\Lt_{m+n}$.
We accordingly relabel the states in terms of their $\Lt_0$ eigenvalues as $|\pm h\rangle \rt |\mp h\rangle$ so that $\Lt_0 |h\rangle = h|h\rangle$, $\Lt_1 |h\rangle =\Lt_{-1}|-h\rangle=0$.  We now have
\bea\label{bgb} V_{h,h'}(z_1,z_2)= z_2^{-2h} \langle h,h'| e^{{1\over z_2} \Lt_{1}}    P e^{{6\over c}\int_{z_1}^{z_2} H_I(z)dz}e^{\Lt_{-1} z_1} | h,h'\rangle
\eea
with
\bea\label{bfa} H_I(z) =  -\big(\Lt_{1}-2z\Lt_0+z^2\Lt_{-1}\big)T(z)~.
\eea

We are interested in computing $ V_{h,h'}(z_1,z_2)$ at large $c$.  There are various limits depending on how $h$, $h'$, and the Lorentzian time separation behave as we take $c$ large.  Before turning to the main case of interest, let us show how to recover a known result  obtained in the limit ${h\over c}, {h'\over c} \rt 0$ with ${hh'\over c}$ and the coordinates $z_{1,2}$ held fixed.  In this regime $T(z)$ can be replaced by its expectation value in the state $|h'\rangle$, so $T(z) = {h'\over z^2}$. Similarly, the $\Lt_n$ appearing in $H_I$   mutually commute amongst themselves in this limit.     We then have
\bea\label{bfaa} {6\over c} \int_{z_1}^{z_2} H_I(z) dz = -{6h'\over c}  \left( {z_2-z_1 \over z_1 z_2} \Lt_1 -2\ln {z_2 \over z_1}\Lt_0+(z_2-z_1)\Lt_{-1}\right)  . \eea
Using this along with
\bea\label{bfab}  e^{-\Lt_{-1} z_1}\Lt_1  e^{\Lt_{-1} z_1} &=& \Lt_1 +2z_1 \Lt_0  +z_1^2 \Lt_{-1} \cr
e^{-\Lt_{-1} z_1}\Lt_0  e^{\Lt_{-1} z_1} &=&  \Lt_0 +z_1 \Lt_{-1} \cr
e^{-\Lt_{-1} z_1}\Lt_{-1}  e^{\Lt_{-1} z_1} &=&\Lt_{-1}  \eea
we obtain
\bea\label{bfac} V_{h,h'}(z)= z^{-2h} \langle h,h'| e^{{1\over z} \Lt_{1}}  e^{\Lt_{-1} }   e^{-{6h'\over c} \left( {z-1 \over  z} \Lt_1 +2(1-{1\over z}-\ln z \Lt_0+(z-{1\over z}-2\ln z \Lt_{-1}\right) } | h,h'\rangle
\eea
where we  now set $z_2=z$, $z_1=1$.   Again, we are allowed to treat $\Lt_n$ in the last factor as mutually commuting, and make the replacements $\Lt_1 \rt 0$, $\Lt_0 \rt h$, $\Lt_{-1} \rt -{2h\over 1-z}$, the latter coming from observing that $\Lt_{-1}$ insertions are obtained differentiating the $c=\infty$ correlator: $\langle h| Pe^{\int_{z_1}^{z_2} \Lt_{-1} dz}\Lt_{-1}|h\rangle = -\p_{z_1}\langle h| Pe^{\int_{z_1}^{z_2} \Lt_{-1} dz}|h\rangle \sim -\p_{z_1} (z_1-z_2)^{-2h}$.       This finally gives the result
\bea\label{bfad} V_{h,h'}(z) =  e^{2{hh'\over c} g_2(1-z) } V^{(c=\infty)}_{h,h'}(z)~,\quad  V^{(c=\infty)}_{h,h'}(z)=   z^{-2h} \langle h| e^{{1\over z} \Lt_{1}}  e^{\Lt_{-1} } |h\rangle \sim (1-z)^{-2h} \eea
where the global stress tensor block is $g_2(1-z)=(1-z)^2 {_2}F_1(2,2,4,1-z) = -12 -6\left( {1+z\over 1-z} \right) \ln z$. The fact that in this limit the global stress tensor block exponentiates was first noted in appendix B of \cite{FitzpatrickKW14}.

Now we turn to the case of primary interest corresponding to large Lorentzian time separations.  We take $c, t\rt \infty $ with ${t\over c}$ fixed.  Here $t$ denotes the time separation between the two $\Oc_h$ operators on the Lorentzian cylinder.  To elucidate this, let's consider the analytic structure of $V_{h,h'}(z_1,z_2)$ in the complex $z_2$ plane, for fixed $z_1$.  There is a branch cut emanating from $z_1$ corresponding to the location of an operator $\Oc_h(z_1)$, as well as a branch cut running between $0$ and $\infty$ corresponding to operators $\Oc_{h'}(0)$ and $\Oc_{h'}(\infty)$.
We are interested in evaluating $V_{h,h'}(z_1,z_2)$ at  late time on the Lorentzian cylinder. The continuation to Lorentzian signature is obtained by taking $z=e^{-i(\phi-t)}$. Hence, taking the points $z_{1,2}$ to be separated by a large Lorentzian time interval corresponds to considering a Wilson line that wraps many times around the branch point at $z=0$.  In particular, if we take it to wrap $N$ times corresponding to $\Delta t = 2\pi N$ then we can write
\bea\label{bl} V_{h,h'}(z_1,z_2)= z_2^{-2h} \langle h,h'| e^{{1\over z_2} \Lt_{1}}  \underbrace{Pe^{{6\over c}\oint_C  H_I(z)dz}\ldots  P e^{{6\over c}\oint_C  H_I(z)dz}}_{\text{N times}}e^{\Lt_{-1} z_1} | h,h'\rangle
\eea
where the contour $C$ goes once counterclockwise around the origin.  For strictly real Lorentzian time both $z_1$ and $z_2$ lie on the unit circle.   The correlator in the $c\rt \infty$ limit is $2\pi$ periodic in $t$ and hence has an infinite number of lightcone singularities.  It will eventually be convenient to regulate these by displacing $z_2$ slightly off the unit circle, corresponding to keeping a small imaginary time component.

We focus now on the light-light limit, where we keep $h$ and $h'$ fixed as $c\rt \infty$. This is the most tractable of the late time limits for the following reason.   What makes (\ref{bl}) difficult to evaluate is the path ordering, which requires us to expand the exponential and compute nested integrals. In the light-light limit, each exponent is suppressed by $1/c$.  At the same time, there are $N\sim c$ exponentials, so combinatoric factors compensate for the $1/c$ suppression of each term.   In this regime each exponential can be expanded to first order, i.e  $\lim_{N \rt \infty} \prod_{i=1}^N e^{a_i\over N} = \lim_{N \rt \infty} \prod_{i=1}^N (1+{a_i\over N})$.   This implies a drastic simplification: since at most one stress tensor appears in the expansion of each exponential, the path orderings are not needed. We simply get $N$ factors of a common exponential,
\bea\label{bm} V_{h,h'}(z_1,z_2)=  z_2^{-2h} \langle h,h'| e^{{1\over z_2} \Lt_{1}} e^{{6N\over c}\oint_C  H_I(z)dz}e^{\Lt_{-1} z_1} | h,h'\rangle .
\eea
Next, we use the standard mode expansion of the stress tensor
\bea\label{bn}
T(z) = \sum_{n} {L'_n \over z^{n+2}} \quad  \Rightarrow \quad  L'_n = {1\over 2\pi i} \oint dz z^{n+1}T(z)~,
\eea
which gives
\bea\label{bo} \oint_C H_I(z)dz = 4\pi i  \Lt \cdot L'.
\eea
The SL(2) dot product is
\bea\label{bp} \Lt \cdot L' = -{1\over 2} \Lt_1 L'_{-1} +\Lt_0 L'_0 -{1\over 2}\Lt_{-1} L'_1~.
\eea
We now have
\bea\label{bo} V_{h,h'}(z_1,z_2)=z_2^{-2h}  \langle h,h'|  e^{{1\over z_2} \Lt_{1} z_2 } e^{{24 \pi i N\over c}\Lt \cdot L'}e^{\Lt_{-1} z_1} | h,h'\rangle.
\eea

Now we want to decompose the states $e^{\Lt_{-1} z_1} | h,h'\rangle$  and $ \langle h,h'|  e^{{1\over z_2} \Lt_{1} z_2 } $ into eigenstates of $(\Lt + L')^2$ and then write $\Lt \cdot L'$ in terms of quadratic Casimirs,
\bea\label{boa}  2\Lt\cdot L' =  (\Lt+L')^2 - \Lt^2-L'^2 \eea
as is familiar from the treatment of spin-orbit coupling in quantum mechanics.  The decomposition of the tensor product is the same problem one encounters in the OPE of two generalized free fields, where one writes $\Oc_h(z_2) \Oc_{h'}(0)  = \sum_{n,k=0}^\infty z_2^{n+k} C_{n,k} \p^k [\Oc_h \Oc_{h'}]_n$.  Here $\p^k [\Oc_h \Oc_{h'}]_n$ is the level $k$ descendant of the double trace quasi-primary operator $[\Oc_h \Oc_{h'}]_n$.  In our case we have
\bea\label{bp} e^{\Lt_{-1} z_1} | h,h'\rangle = \sum_{n,k=0}^\infty z_1^{n+k} C_{n,k} |h+h'+n\rangle_k~. \eea
The state $ |h+h'+n\rangle_k$ is the unit normalized, level $k$ descendant state
\bea\label{bq}  |h+h'+n\rangle_k =  {(\Lt_{-1}+L'_{-1})^k |h+h'+n\rangle \over \sqrt{k!(2h+2h'+2n)_k}}  \eea
where $ |h+h'+n\rangle$ denotes a primary state with respect to the ``total" SL(2) generators.  The coefficients are given as
\bea\label{br}  C_{n,k} =  {(-1)^n  \over \sqrt{n! k!}} \left(  { (2h)_n(2h')_n (2h+n)_k^2\over (2h+2h'+n-1)_n(2h+2h'+2n)_k}\right)^{1/2}.   \eea
It is now straightforward to evaluate (\ref{bo}); however, it is even simpler to use the connection to the generalized free field problem.  In particular, if we take the $c\rt \infty $ limit, then $V_{h,h'}(z_1,z_2)$ is given by the exchange of the identity operator $\Oc_h \Oc_h \rt 1 \rt \Oc_{h'}\Oc_{h'}$, and our decomposition problem is equivalent to reexpressing this in the crossed channel, $\Oc_h \Oc_{h'} \rt [\Oc_h \Oc_{h'}]_{n} \rt \Oc_h \Oc_{h'}$.  The solution is \cite{FitzpatrickK11}
\bea\label{bs} && z_2^{-2h}  \langle h,h'|  e^{{1\over z_2} \Lt_{1} z_2 }e^{\Lt_{-1} z_1} | h,h'\rangle \cr &&= \sum_{n=0}^\infty  {(2h)_n (2h')_n \over n! (2h+2h'+n-1)_n} \left(z_1 \over z_2\right)^n {_2}F_1(2h+n,2h+n,2h+2h'+2n,z_1 z_2^{-1} ). \cr &&
\eea
This is the expansion of the identity operator exchange in terms of crossed channel global conformal blocks.
Given this result we can easily modify it to compute (\ref{bo}).  Each state in the decomposition is an eigenstate of $\Lt \cdot L'$ with eigenvalue given by the quadratic Casimirs in (\ref{boa}), and we therefore have
\bea\label{bs} V_{h,h'}(t) = \sum_{n=0}^\infty  {(2h)_n (2h')_n \over n! (2h+2h'+n-1)_n} r^n {_2}F_1(2h+n,2h+n,2h+2h'+2n,r )e^{-i\frac{\gamma_n}{2} t } \cr &&
\eea
with
\bea\label{bu} \gamma_n = -{12\over c} \Big[C_2(h+h'+n)-C_2(h)-C_2(h')   \Big] \eea
where  $C_2(h) = h(h-1)$.   We have written $t = 2\pi N$, and we have also taken ${z_1 \over z_2}=r$, with $0<r < 1$.  As mentioned above, taking $r<1$ corresponds to giving an imaginary part to the final Lorentzian time, which regulates the lightcone singularity when $z_2$ becomes null separated from $z_1$ on the Lorentzian cylinder.  Also, note that  $V_{h,h'}(t)$ should be thought of as the ``stripped correlator", which does not include the phase factor $e^{-iht}$ associated with the $\Oc_h$ operator on the cylinder. The expression (\ref{bs}) for the late time block is the main result of this paper.

Recall that we have only computed one chiral half of the full stress tensor contribution to the correlator; with the chosen operator locations, including the other chiral half just corresponds to squaring the above result. Upon doing so, $\gamma_n$ would then be replaced by $\gamma_n +\gamma_{\nb}$.  In the next section we discuss the relation between this result and the anomalous dimensions of double trace operators of schematic form $\Oc \p^n \pb^{\nb} \Oc'$.

To conclude this section, in figures 1-3 we give some representative plots of $V_{h,h'}(t)$.  We set $h=3.23$, $h'=4.91$, $r=0.9$, and plot $\ln\left( {V_{h,h'}(t) \over V_{h,h'}(0)}\right)$ over various time windows.   At early times we see a decaying behavior similar to what one observes for correlators in a black hole background, where the decay rate is set by the temperature.  Here, however, the decay rate is non-universal, as it depends on the value of the regulator $r$, which is not surprising given that we do not expect to be seeing thermalization in this regime.  At later time we see the characteristic behavior associated with adding together a large but finite number of  phase factors.  In particular, we can see erratic behavior together with signs of recurrences.  Again, while this is the sort of behavior we would expect to see when computing correlators in a thermal system with a discrete spectrum, we have no reason to expect that the similarity is particularly meaningful given that we are far from a thermal regime.
\begin{figure}[h]
  \flushleft
  \begin{minipage}[b]{0.6\textwidth}
    \centering\includegraphics[width=1.4\textwidth]{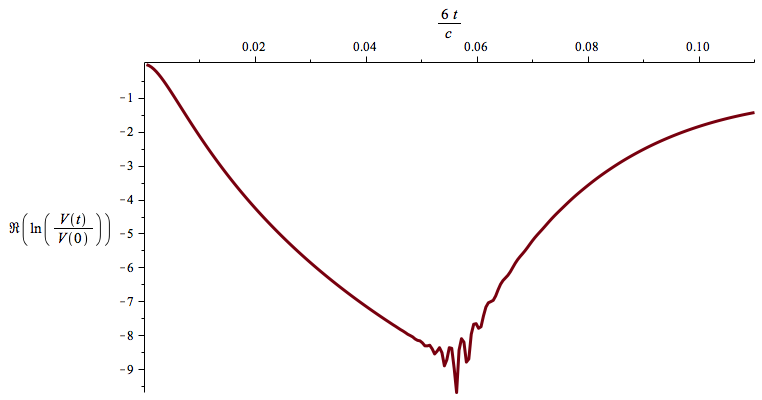}
  \end{minipage}
  \caption{ Plot of Wilson line correlator at early time. We plot the real part of $\ln \left( V_{hh'}(t) \over V_{hh'(0)}\right)$ versus ${6t\over c}$, with the parameter choices $h=3.23$, $h'=4.91$, $r=0.9$. }
  \label{fig:fa}
\end{figure}
\begin{figure}[h]
  \flushleft
  \begin{minipage}[b]{0.6\textwidth}
    \centering\includegraphics[width=1.4\textwidth]{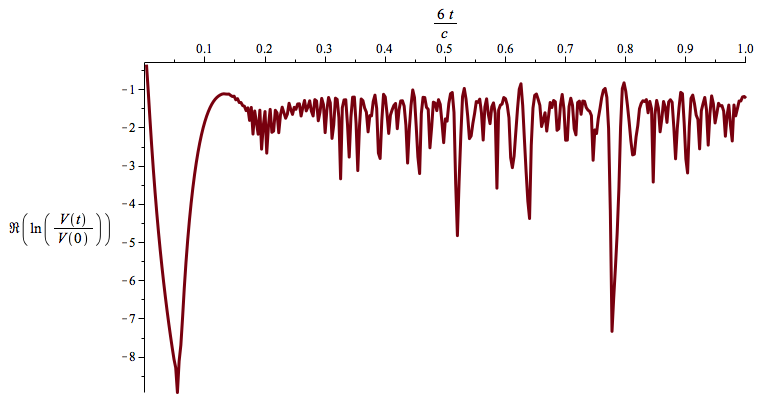}
  \end{minipage}
  \caption{ Same parameters as  in Fig. \ref{fig:fa} but now over larger time range  }
  \label{fig:fb}
\end{figure}
%
\begin{figure}[h]
  \flushleft
  \begin{minipage}[b]{0.6\textwidth}
    \centering\includegraphics[width=1.4\textwidth]{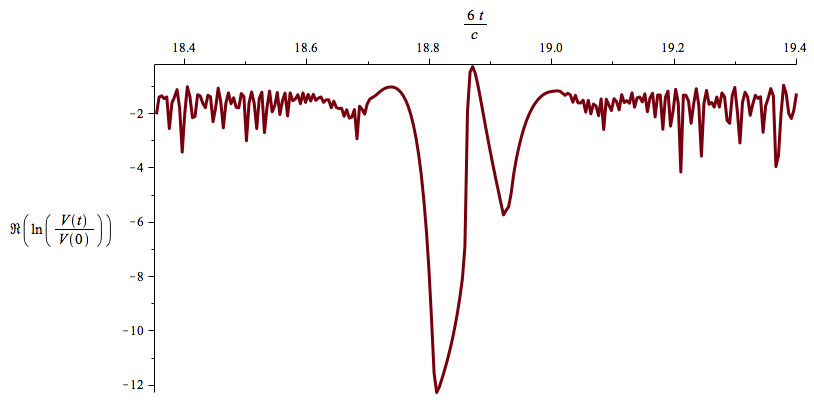}
  \end{minipage}
  \caption{ Same parameters as  in Fig. \ref{fig:fa} but now at late time }
  \label{fig:fc}
\end{figure}
\newpage

\section{Anomalous dimensions from OPE inversion }

The recently-derived Lorentzian inversion formula  \cite{OPEInversion,Simmons-DuffinSW17} is an efficient tool for extracting CFT data --- i.e. OPE coefficients and anomalous dimensions ---  from correlation functions.  In this section we will review its use as applied to the problem at hand, namely stress tensor exchange contributions to $d=2$ four-point functions.  This is an instructive example to work through. We note that this is just a special case of the general analysis performed in \cite{LiuPRS18}, although by focusing on this one case we are able to give results for all spins and twists that are more explicit.

Given scalar operators $\Oc_{1,2}$   of dimension $\Delta_1$ and $\Delta_2$, we consider the four-point function
\begin{equation}
\braket{\mathcal{O}_1(x_1)\mathcal{O}_2(x_2) \mathcal{O}_2(x_3) \mathcal{O}_1(x_4)}
 \equiv (x_{12}^2 x_{34}^2)^{-{\Delta^+_{12}\over 2}}
\left( \frac{x_{14}^2}{x_{24}^2} \frac{x_{14}^2}{x_{13}^2}\right)^{-{\Delta^-_{12}\over 2}} G(z, \bar{z}),
\end{equation}
with $x_{ab}^2 = (x_a-x_b)^2$.
We are using the notation
\bea\label{ya}\Delta^{\pm}_{12} = \Delta_1\pm \Delta_2~,
\eea
and the conformal  cross ratios are
\bea\label{yb}
z\zb =  \frac{x_{12}^2 x_{34}^2}{x_{13}^2 x_{24}^2}~,\quad (1-z)(1-\zb) = {x_{23}^2 x_{14}^2 \over x_{13}^2 x_{24}^2}~.
\eea
We then use conformal invariance to send three points to specified locations and write
\bea\label{ye}
\braket{\mathcal{O}_1(0)\mathcal{O}_2(z, \bar{z}) \mathcal{O}_2(1) \mathcal{O}_1(\infty) } = (z\zb)^{-{\Delta^+_{12}\over 2}}  G(z, \bar{z})~.
\eea
The small $(z,\zb)$ expansion of $ G(z, \bar{z}) $ contains information about the operators that appear in the $\Oc_1 \Oc_2$ OPE, in particular their dimensions and OPE coefficients. The dimensions and OPE coefficients of primary operators are read off from the conformal block expansion of $G(z,\bar{z})$, while conformal invariance fixes the data concerning descendants.

We are working in the context of a theory that is a small perturbation around so-called generalized free fields (equivalently, Mean Field Theory), as appropriate for matching to a weakly coupled theory of gravity in AdS$_3$. Bulk tree level  exchange diagrams contributing to the four-point function can be labeled as s, t, or u channel diagrams in the standard fashion.   What the inversion formula allows us to do is to efficiently extract s-channel OPE data (associated to the $\Oc_1 \Oc_2$ OPE) from the t and u channel diagrams.  More precisely, we can extract this data for  operators whose spin is larger than some critical value, where the critical value is set by the spin of the exchanged fields in AdS$_3$. In our case, we will only have t-channel diagrams to the order we are working, since we assume there is no bulk vertex directly coupling $\Oc_1$ and $\Oc_2$.  A very powerful fact is that we will not need the full t-channel exchange Witten diagram in order to extract the s-channel CFT data.  The Witten diagram corresponds to the exchange of a stress tensor block along with double trace blocks, but the latter can be shown to give zero when plugged into the inversion formula \cite{OPEInversion}.  This is a great advantage, since the full Witten diagram is a somewhat complicated beast, while the conformal block is readily available.

The CFT data of interest is encoded in the poles and residues of the function $c(\Delta,J)$, which is obtained from the correlator via the inversion formula
\begin{align*}
c(\Delta,J) = \frac{\kappa}{4} \int_0^1 & \frac{dz}{z^2} \frac{d\bar{z}}{\bar{z}^2}
{g_{ J+1,\Delta-1}(z, \bar{z})\over \big((1-z)(1-\zb)\big)^{\Delta^-_{12}}}
\dDisc\left[
 G(z,\bar{z}) \right].
\numberthis
\end{align*}
The normalization factor is
\begin{equation}
\kappa = \frac{\Gamma^2\left({\Delta+J+\Delta^-_{12}\over 2} \right) \Gamma^2 \left({\Delta+J-\Delta^-_{12}\over 2}  \right)}{2 \pi^2 \Gamma(\Delta+J-1)\Gamma(\Delta+J)},
\end{equation}
and the 2d conformal blocks are
\begin{equation}
g_{\Delta,J}(z, \bar{z}) = \frac{k_{\Delta-J}(z) k_{\Delta+J}(\bar{z}) + k_{\Delta+J}(z) k_{\Delta-J}(\bar{z})}{1 + \delta_{J,0}},
\end{equation}
with
\begin{equation}
k_\beta(z) = z^{\beta/2} \F \left({\beta -\Delta^-_{12}\over 2}, {\beta -\Delta^-_{12}\over 2}, \beta, z \right).
\end{equation}
The relevant formula for the double discontinuity, dDisc$[G(z,\zb)]$, appears below. {The function $c(\Delta,J)$ has poles in real $\Delta$ at the location of primaries $\mathcal{O}_p$ exchanged in the s channel. Near $\Delta_p$,
\begin{equation}
c(\Delta,J) \sim -\frac{C_{\Oc_1 \Oc_2 \Oc_p}^2}{\Delta-\Delta_p},
\label{ycc}
\end{equation}
so that $-\text{Res}(c(\Delta,J))_{\Delta = \Delta_p}$ is the square of the OPE coefficient.

Since we are in d=2 the conformal group factorizes, and it is useful to make this explicit.   Operators are labelled by scaling dimensions $(h,\hb)$, related to the dimension and spin by $\Delta=h+\hb$, $J=|h-\hb|$.  Is it convenient to assume $h\geq \hb$ for the exchanged operators so that $J=h-\hb$; there is no loss of information here, since parity invariance implies that the CFT data is invariant under $h\leftrightarrow \hb$.   We thus write $c(h,\hb)=c(\hb,h)$ with
\begin{align*}
c(h,\hb) = \frac{\kappa}{2 } \int_0^1 & \frac{dz}{z^2} \frac{d\bar{z}}{\bar{z}^2}
{g_h(z)g_{1-\hb}(\zb) \over \big((1-z)(1-\zb)\big)^{2h^-_{12}}}
\dDisc\left[
 G(z,\bar{z}) \right],
\numberthis
\label{InversionFormula}
\end{align*}
where
\bea\label{yc} g_h(z) = z^h \F (h -h^-_{12}, h-h^-_{12}, 2h, z) \eea
and $h^\pm_{12}= h_1 \pm h_2$.

Now we consider the contribution due to a single conformal block in the t-channel with quantum numbers $(H,\Hb)$,
\bea\label{yd}  \braket{\mathcal{O}_1(0)\mathcal{O}_2(z, \bar{z}) \mathcal{O}_2(1) \mathcal{O}_1(\infty) } = {1\over [(1-z)(1-\zb)]^{2h_2} } f_H(1-z)f_{\Hb}(1-\zb)~,  \eea
where we have suppressed the OPE coefficient, and we  note that by parity there will also be a contribution from an exchange with $H\leftrightarrow \Hb$.  Here  $f_H(1-z) = (1-z)^H \F(H,H,2H,1-z)$ is a t-channel conformal block.  Comparing to (\ref{ye}) we have
\bea\label{yf} G(z,\zb) =  {(z\zb)^{h^+_{12}} \over [(1-z)(1-\zb)]^{2h_2} } f_H(1-z)f_{\Hb}(1-\zb)~.  \eea
The inversion formula involves the double discontinuity $\dDisc\left[
 G(z,\bar{z}) \right]$, which is defined in terms of the analytic continuation of $\zb$ around the branch cut emanating from $\zb=1$
\begin{equation}
\dDisc(G) = \cos(2\pi h^-_{12})G-\frac{1}{2}\left( e^{2 \pi i h^- _{12}} G^\circlearrowright+e^{-2 \pi i h^-_{12}  } G^\circlearrowleft \right),
\end{equation}
where $G^\circlearrowright, G^\circlearrowleft$ are continuations $1-\bar{z} \rightarrow e^{-2 \pi i }(1-\bar{z}),~e^{2 \pi i }(1-\bar{z})$ respectively. For our purposes we just need
\begin{align*}
\mathcal{D} \equiv \frac{\dDisc((1-\bar{z})^{-2h_2})}
{(1-\bar{z})^{-2h_2}}
 &= 2 \sin(\pi(2h_1)) \sin(\pi(2h_2))
\\
&= 2\frac{\pi}{\Gamma(2h_1)\Gamma(1-2h_1)}\frac{\pi}{\Gamma(2h_2)\Gamma(1-2h_2)}.
\end{align*}
Also, in present notation we have
\bea\label{yg} \kappa = {\Gamma^2(h+h^-_{12})\Gamma^2(h-h^-_{12}) \over 2\pi^2 \Gamma(2h-1)\Gamma(2h)}~. \eea
It follows that   the contribution to $c(h,\hb)$ from the t-channel exchange of $(H,\Hb)$ is
\begin{equation}
c(h,\hb) = \frac{\mathcal{D} \kappa}{2} \mathcal{I}_H(h)\mathcal{I}_{\Hb}(1-\bar{h}),
\label{CompactIdentityExpression}
\end{equation}
where
\begin{equation}
 \mathcal{I}_H(h) \equiv \int_0^1 {dz\over z^2} {z^{h^+_{12}} \over (1-z)^{2h_1}} g_h(z) f_H(1-z).
\label{CaronHuot4.7}
\end{equation}
To remind ourselves of its $\zb$ origin we write $\mathcal{I}_{\Hb}(1-\hb) \equiv \int_0^1 {d\zb\over \zb^2} {\zb^{h^+_{12}} \over (1-\zb)^{2h_1}} g_{1-\hb}(\zb) f_{\Hb}(1-\zb)$.

\subsection{Identity exchange}

At the level of free field theory in the bulk we only have a contribution from a disconnected diagram, which corresponds to exchange of the CFT identity operator, $(H,\Hb)=(0,0)$.  To evaluate this we need the integral
\bea\label{yh} \Ic_0(h) &=&  \int_0^1 {dz\over z^2} {z^{h^+_{12}} \over (1-z)^{2h_1}} g_h(z) \cr
& =& {\Gamma(1-2h_1)\Gamma(1-2h_2)\Gamma(2h)\Gamma(h+h^+_{12}-1)\over \Gamma(h-h^-_{12})\Gamma(h+h^-_{12})\Gamma(h-h^+_{12}+1 )}
\eea
This result may be obtained by using the Euler integral representation for the hypergeometric function and then changing variables to decouple the two integrals (see \cite{AldayC17}).  We then note that $\Ic_0(1-\hb)$ has simple poles at $\hb = h^+_{12}+\nb$, ($\nb=0,1,2,\ldots$),\footnote{There are also some other ``spurious poles" which end up giving no contribution, and we ignore them here;  see \cite{OPEInversion,Simmons-DuffinSW17} for details.}
%
%
%
\bea\label{yi} \Ic_0(1-\hb) \sim -{1\over \nb!}{(2h_1)_{\nb} (2h_2)_{\nb}\over (2h^+_{12}+\nb-1)_{\nb}}{1\over \hb-h^+_{12}-\nb}  \quad {\rm as}\quad  \hb \rt h^+_{12}+\nb
\eea
where $(a)_{\nb} = {\Gamma(a+\nb)\over \Gamma(a)}$ is the Pochhammer symbol, and we used the identity $ {\Gamma(x+\nb)\over \Gamma(x) } = (-1)^{\nb}{\Gamma(1-x)\over \Gamma(1-x-\nb)}$.

At the pole location we have $h=\hb+J = h^+_{12}+\nb +J$, which we write as $h=h^+_{12}+n$ with $n=\nb+J$.  At such a pole we then evaluate
\bea\label{yj} \mathcal{D} \kappa \Ic_0(h) = {1\over n!}{(2h_1)_{n} (2h_2)_{n}\over (2h^+_{12}+n-1)_{n}}~,\quad h=h^+_{12}+n \eea
We therefore find  simple poles in $c(h,\hb)$  according to
\bea\label{yk} c(h,\hb) \sim -\frac{1}{2}{ C_{n,\nb} \over \hb-h^+_{12}-\nb}\quad {\rm as} \quad \hb \rt h^+_{12}+\nb~,\quad h= h^+_{12}+n~,\eea
with
\bea\label{yl}  C_{ n,\nb} = \left[  {1\over n!}{(2h_1)_{n} (2h_2)_{n}\over (2h^+_{12}+n-1)_{n}} \right] \left[{1\over \nb!}{(2h_1)_{\nb} (2h_2)_{\nb}\over (2h^+_{12}+\nb-1)_{\nb}}\right] ~.\eea
The factor of $1/2$ in \ref{yk} arises from using $h,\bar{h}$ rather than $\Delta,J$. This identifies $ C_{ n,\nb}$ as the squared OPE coefficients of primary operators appearing in the $\Oc_1 \Oc_2$ OPE in the generalized free field limit.
One checks that these are the correct OPE coefficients  by verifying (see (\ref{yf}) the identity
\bea\label{ym} {(z\zb)^{h^+_{12}} \over [(1-z)(1-\zb)]^{2h_2} }  = \sum_{n,\nb=0}^\infty  C_{ n,\nb} g_{h^+_{12}+n}(z) g_{h^+_{12}+\nb}(\zb)~. \eea
That is, this gives the expansion of the vacuum exchange contribution  $\Oc_1 \Oc_1 \rt I \rt \Oc_2 \Oc_2$ in terms of crossed channel exchanges, $\Oc_1 \Oc_2 \rt [\Oc_1 \Oc_2]_{n,\nb} \rt \Oc_1 \Oc_2$.

\subsection{Anomalous dimensions from graviton exchange}

We now include graviton exchange in the t-channel. At large $c$  we can relate this to $1/c$ corrections to the anomalous dimensions.  If the previous poles at $\hb =h^+_{12}+\nb$ are shifted to $\hb=h^+_{12}+\nb+{\gamma\over 2}$ where $\gamma \sim {\cal O}(1/c)$ then the pole term will appear in $1/c$ perturbation theory as
\bea\label{yo} c(h,\hb) \sim -{1\over 2} {C_{n,\nb} \over \hb-h^+_{12}-\nb-{\gamma\over 2}} =  -{1\over 2} {C_{n,\nb} \over \hb-h^+_{12}-\nb} -{1\over 2} {C_{n,\nb} \over (\hb-h^+_{12}-\nb)^2}{\gamma\over 2}+ {\cal O}(1/c^2)~. \eea
There is also a $1/c$ contribution from corrections to $C_{n,\nb}$ that we have suppressed.  Hence anomalous dimensions are extracted from the coefficient of the double pole.

As noted above, we really just need to explicitly compute the contributions from the stress tensor rather than the full Witten diagram.  So, adding this to the identity exchange we have
\bea\label{yn} c(h,\hb) = {\mathcal{D} \kappa\over 2} \Big( \Ic_0(h)\Ic_0(1-\hb) + {2h_1 h_2 \over c} \Ic_0(h) \Ic_2(1-\hb) +  {2h_1 h_2 \over c} \Ic_2(h) \Ic_0(1-\hb)\Big)~.\eea
The factors of ${2h_1 h_2 \over c}$ are understood as follows.  The usual OPE expression is $T(z) \Oc(h) \sim {h\over z^2} \Oc(0) +\ldots$, identifying $h$ as the OPE coefficient.   However, this definition of the stress tensor has two-point function   $\langle T(z)T(0)\rangle = {c\over 2z^4}$, so in (\ref{yn}) we need to divide by $c/2$ to describe the exchange of a properly normalized operator.

Looking back at the identity exchange computation, simple poles arose from the $\zb \rt 0$ region of integration according to   $\int_0^1 \! d\zb \zb^{\hb_{\nb}-\hb-1} \sim -{1\over \hb-\hb_{\nb}  }$, with $\hb_{\nb}= h^+_{12}+\nb$.   Double poles arise from the presence of a logarithm,
\bea\label{yna} \int_0^1 \! d\zb \zb^{\hb_{\nb}-\hb-1} \ln \zb \sim -{1\over (\hb-\hb_{\nb})^2 }~, \eea
so that, as will use below, the coefficient of the double pole is the same as the coefficient of the simple pole for the integral with no $\ln \zb$ insertion.
Such  a logarithm comes from $f_2(1-\zb)$ appearing in $\Ic_2(1-\hb)$,
\bea\label{yo} f_2(1-\zb) = (1-\zb)^2 \F(2,2,4,1-\zb) = -12- 6\left({2\over 1-\zb}-1\right)\ln \zb~. \eea
The upshot is that the anomalous dimension comes from the middle term  in (\ref{yn}), and we can further omit the $-12$ term in $f_2(1-\zb)$, in which case
\bea\label{yp} \Ic_2(1-\hb) =  -6\int_0^1 {d\zb\over \zb^2} {\zb^{h^+_{12}} \over (1-\zb)^{2h_1}} g_{1-\hb}(\zb)\left({2\over 1-\zb}-1\right)\ln \zb~. \eea
Now, as we noted above, the coefficients of the double poles in $\Ic_2(1-\hb)$  are the same as the coefficients of the simple poles after omitting $\ln \zb$ from  the integrand.  The integral with coefficient $(-1)$ is simply $\Ic_0(1-\hb)$ and the corresponding simple poles are written in (\ref{yi}).  For the remaining integral we find poles
\bea\label{yq}
 && \int_0^1 {d\zb\over \zb^2} {\zb^{h^+_{12}} \over (1-\zb)^{2h_1}} g_{1-\hb }(\zb)\left({2\over 1-\zb}\right) \sim -{1\over \nb!}{(2h_1)_{\nb} (2h_2)_{\nb}\over (2h^+_{12}+\nb-1)_{\nb}}\left[ (2h_1+n)(2h_2+n)-n \over 2h_1 h_2  \right] {1\over \bar{h}-h^+_{12}-n} \cr && \eea
 This result is obtained by expanding the integrand around $\zb=0$, and extracting the coefficients using  some guesswork and checking.
This gives  the double poles
\bea\label{yr} \Ic_2(1-\hb) \sim  {1\over \nb!}{(2h_1)_{\nb} (2h_2)_{\nb}\over (2h^+_{12}+\nb-1)_{\nb}} \left[  {6\over 2 h_1 h_2 }\Big( C_2(h_{12}^++n)-C_2(h_1)-C_2(h_2) \Big) \right]  {1\over (\hb-h^+_{12}-\nb)^2}.  \cr && \eea
where the SL(2) quadratic Casimir $C_2(h)=h(h-1)$ has appeared.
Using this in (\ref{yn})  we have
\bea\label{ys} c(h,\hb)  \sim -{1\over 2} C_{n,\nb} \left( {1\over \hb-h_{12}^+-\nb} +{\gamma_{\nb}/2 \over  (\hb-h^+_{12}-\nb)^2}\right)~,\eea
with
\bea\label{yt} \gamma_{\nb}= -{12 \over c} \Big( C_2(h_{12}^++\nb)-C_2(h_1)-C_2(h_2) \Big) ~.\eea
Recall that we assumed $n\geq \nb$, but if we relax this condition then $\nb$ in (\ref{yt}) should be replaced by min$(n,\nb)$.
$\gamma_{{\rm min}(n,\nb)}$ gives the anomalous dimension at order $1/c$ of the operator of schematic form $\Oc_1 \p^n \pb^{\nb} \Oc_2$.  That is, this operator has dimension $\Delta = h_{12}^+ +n+\nb+ \gamma_{{\rm min}(n,\nb)}$ and spin $J=|n-\nb|$.   Noting that the twist in the generalized free limit  is $\tau = \Delta-J = h_{12}^+ + 2~{\rm min}(n,\nb)$ we have that when expressed in terms of $(\tau,J)$ the anomalous dimension depends on the twist but not the spin.

\subsection{Comparison to Wilson line}

 Comparing to our Wilson line computation, we see that the anomalous dimensions $\gamma_{n,\nb}$ has the same form as the coefficients $\gamma_n$ obtained in the Wilson line computation, except for the replacement of ${\rm min}(n,\nb)$ by $n$.  The difference is that in this section we are expanding a full correlator in the crossed channel, while the Wilson line just gives the Virasoro vacuum block contribution.

To further illustrate the distinction we carry out the following exercise.  Suppose we pretend that the correlator is given by the t-channel exchange of the identity operator and the global stress tensor block (this corresponds to expanding the Virasoro block contribution to first order in $1/c$).  We  just focus on one chiral half of the correlator in what follows.    This gives
\bea\label{yu} G(z)& =& {z^{h^+_{12}}\over (1-z)^{2 h_2}} \left(1+ {2h_1h_2 \over c} z^2F(2,2,4,1-z)\right) \cr& =&{z^{h^+_{12} }\over (1-z)^{2 h_2}} \left(1+ {2h_1h_2 \over c} \big(-12-6({2\over 1-z}-1) \ln z \big)   \right)~.\eea
We then try to expand this in the s-channel in terms of double trace operators with $1/c$ corrected dimensions and OPE coefficients,
\bea\label{yv}G(z) = \sum_{n=0}^\infty (C_n+\delta C_n) g_{h^+_{12}+{\gamma_n\over 2}}(z)~,\eea
with $C_n = {(2h_1)_n (2h_2)_n \over n! (2h_1+2h_2+n-1)_n}$.  The $\gamma_n$ are extracted by comparing the $\ln z$ terms on two sides after expanding to order $1/c$ and recalling that $\gamma_n \sim O(1/c)$.
\bea\label{yw}  {z^{h_1+h_2}\over (1-z)^{2 h_2}}{2h_1h_2 \over c} \big(-6({2\over 1-z}-1)  \big)\ln z={1\over 2} \sum_{n=0}^\infty C_n  \gamma_n g_{h_1+h_2+n}(z) \ln z~. \eea
One can check that this is obeyed for
\bea\label{yx} \gamma_n = -{12\over c}  \big( C_2(h^+_{12}+n)-C_2(h_1)-C_2(h_2) \big) \eea
which is the result found in the Wilson line computation.   We can continue to match the non-$\ln z$ terms to fix $\delta C_n$, at which point we have succeeded in writing $G(z)$ in the form of a block expansion in the s-channel, (\ref{yv}).   Including the $\zb$ contributions would simply yield the absolute square, and anomalous dimensions $\gamma_{n,\nb} = \gamma_n+\gamma_{\nb}$.  However, since we omitted double trace exchanges in the t-channel, even though these are known to be present in a full Witten diagram, one might expect there is  something sick about this result.  For example, the  spectrum appearing in the s-channel includes operators with non-integer spin, since $J= |h-\hb| = {1\over 2} |\gamma_n-\gamma_{\nb}|$ is not an integer in general.  So from this point of view, the double trace exchanges in the t-channel are required to maintain an operator spectrum with purely integer spins. Another way to understand $\gamma_n$ is as anomalous dimensions at large spin. Taking the lightcone limit $\bar{z} \rightarrow 1$,  the operator with minimal twist $\tau_{min} = 0$, the stress tensor, dominates. A familiar result from the lightcone bootstrap is that the leading dependence of the double-twist anomalous dimensions on spin $J$ is $1/J^{\tau_{min}}$, and so the large-spin anomalous dimensions are spin-independent. As this data is analytic in $J$, the anomalous dimensions must take their large-spin value for all spins.  The crossed channel expansion in the lightcone limit is dominated by states with $\nb \gg n$, so that $n$ is the twist, and $n={\rm min}(n,\nb)$.  This  explains why $\gamma_{n,\bar{n}}$ takes the same form as $\gamma_n$.
 Also, this computation shows very clearly how the Wilson line in the limit we have considered corresponds to summing up the single stress tensor exchanges into a form in which
$\gamma_n$  appears exponentiated.

\section{Comments}

We conclude with a few comments.  The main result of this paper is an expression for the late time Wilson line, obtained in the limit $c,t \rt \infty$ with $t/c$ fixed.  This was achieved in the light-light limit, where $h$ and $h'$ are held fixed.  The most obvious challenge for the future is to extend this to the heavy-light limit, in which $h'/c$ is held fixed, which would allow one to make contact with the black hole related issues discussed in the introduction.  We might  hope to gain analytical insight into the numerical results of \cite{ChenHKL17}, which indicate a universal $1/t^{3/2}$ falloff at late times for these blocks.     A step in this direction might be to systematically understand $1/c$ corrections to the light-light limit.   We focused here on the vacuum Virasoro block, but the Wilson line construction is readily generalized to describe non-vacuum blocks as well.   The starting point is a Wilson line network with trivalent vertices. \cite{Bhatta:2016hpz,Besken:2016ooo}. It would be interesting to employ the methods used here to understand the late time behavior of these blocks.  We also note that a closely related approach to Virasoro blocks in the large $c$ limit is based on the ``geometric action" for the Virasoro group \cite{Alekseev:1988ce}.  This was recently considered in \cite{Cotler:2018zff}, along with its appearance from 3D gravity in the Chern-Simons formulation.

Applying familiar bootstrap techniques to the Virasoro-block decomposition of correlators may provide further insight. In this work, we found that the anomalous dimensions $\gamma_{n,\bar{n}}$ due to graviton exchange differed from the corrections to the energy $\gamma_n + \gamma_{\bar{n}}$ in the Virasoro block computation essentially due to the tower of double-trace operators present in Witten diagrams. At order $\mathcal{O}(1/c^0)$, the way the double-trace operators are encoded in the Virasoro-block decomposition is clear, as Virasoro blocks reduce to global blocks in the large-$c$ limit. The role of double-trace operators beyond leading order has been studied for the global block decomposition using the Lorentzian inversion formula \cite{AldayC17} but is less well-understood in the Virasoro block case. One can investigate this by expanding Virasoro blocks in terms of global blocks and working order by order in $1/c$. This procedure computes contributions from multi-stress-tensor blocks in a systematic way, while making use of global conformal symmetry alone would not allow this convenient packaging of gravitational data.

\section*{Acknowledgements}

We thank Mert Besken, Alejandra Castro, Ashwin Hedge, Eric Perlmutter, David Simmons-Duffin, and the participants of the Simons Collaboration on the Non-perturbative Bootstrap workshop held in Caltech in 2018 for useful discussions. P.K. is supported in part by NSF grant PHY-1313986.



\bibliographystyle{ssg}
\bibliography{refs}

\end{document}